# The Lokahi Prototype: Toward the automatic Extraction of Entity Relationship Models from Text

Michael Kaufmann

Lucerne University of Applied Sciences and Arts
School of Information Technology
Suurstoffi 41, 6343 Rotkreuz, Switzerland
m.kaufmann@hslu.ch

## Abstract

Entity relationship extraction envisions the automatic generation of semantic data models from collections of text, by automatic recognition of entities, by association of entities to form relationships, and by classifying these instances to assign them to entity sets (or classes) and relationship sets (or associations). As a first step in this direction, the Lokahi prototype can extract entities based on the TF*IDF measure, and generate semantic relationships based on document-level co-occurrence statistics, for example with likelihood ratios and pointwise mutual information. This paper presents results of an explorative, prototypical, qualitative and synthetic research, summarizes insights from two research projects and, based on this, indicates an outline for further research in the field of entity relationship extraction from text.

## Introduction

With the data explosion we are currently facing, tools to provide an overview are needed. Knowledge extraction techniques could support humans to keep track of important information. If a knowledge management system could extract semantic structure from text automatically, it became possible to automate the task of classifying and ordering data and documents. For example, automatic tagging of e-mails and automatic linking of tags could help alleviate the problem of the e-mail flood. Also, automatic extraction of knowledge networks can help explorative analysis of unstructured data, for example in the field of social media mining. Therefore, in this paper a framework for knowledge extraction based on simplified entity relationship models is presented. A research prototype is described that exemplifies a first step in this direction, and its method for entity extraction and relationship extraction is explained. The paper concludes with insights from this explorative synthesis, and an outline of research questions to achieve the vision of automatically deriving entity relationship models from text.

## Background

The network as a meta-structure of knowledge representation has been postulated for half a century. Instances are Semantic Networks (Quillian, 1967), Conceptual Graphs (Sowa, 1976), Entity-relationship models (Chen, 1976), Concept Maps (Novak & Gowin, 1984), Topic Maps (Rath & Pepper, 1999) and the semantic web, all of which overlap in basic principles but differ in application orientation. Semantic networks serve the knowledge representation for artificial intelligence; conceptual graphs have been developed for use in database systems; Concept maps were used for university didactics; and Topic Maps serve the exchange of metadata via XML (XTM); Semantic Web Technology (RDF) is intended for machine-to-machine knowledge exchange and reasoning

Instead of manually encoding and externalizing knowledge, ontology learning is a technique to automatically infer knowledge networks from data (Maedche & Staab, 2001, Alani et al., 2003). There exist approaches to extract knowledge networks from data. For example, (Böhm, Heyer, Quasthoff, & Wolff, 2002) generated topic maps from text using window-based co-occurrences. (Villalon & Calvo, 2009) used a syntactical approach, analyzing grammatical structures in sentences to induce concept maps.

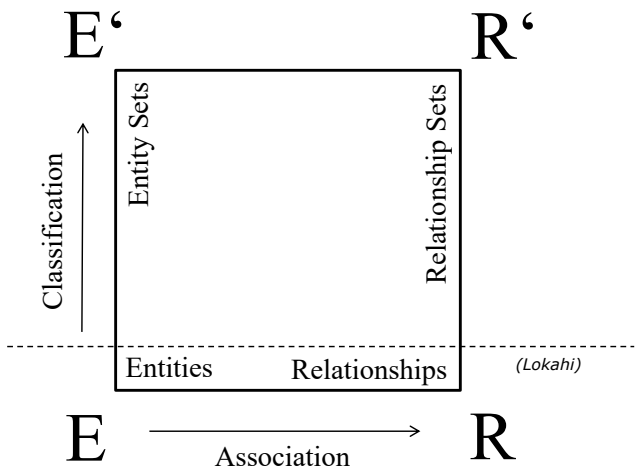

*Figure 1: Vision: a framework for knowledge extraction*

## A Framework for Knowledge Extraction

Figure 1 presents an abstract framework for knowledge extraction. The vision is to automatically infer entity-relationship models in the sense of (Chen, 1976) directly from text. Following Chen (1976), a simplified Entity Relationship (ER) knowledge network can be defined as a quadruple ER = (E, R, E', R'), defined by the following components:

1.) A set of symbols E ⊂ Σ* which are named entities as a subset of arbitrary strings. According to Chen, "An entity is a thing which can distinctly identified. A specific person, company, or event is an example of an entity."

2.) A set R ⊂ E×E of binary relationships between entities, r. Chen wrote that "A relationship is an association between entities." (Chen 1976)

3.) A class E' ⊂ E of entity sets that group or cluster similar entities. Elements of E' are classes with an element relation: ∈ ⊂ E×E'

4.) A set R' ⊂ E of relationship sets that group or cluster similar relationships. This extends the element relation to ∈ ⊂ R×R'

This definition of ER models is simplified in the sense that there are no properties, entities are identified with their name, and relationships can only be binary. Also, it is a purely syntactical approach fitted for extraction from text, where all labels, even entity sets and relationship sets, are named entities, that is, entities identified with their syntactical representation in form of their name. This reduces the task of entity and relationship classification to finding truth tables for the element relation for named entities.

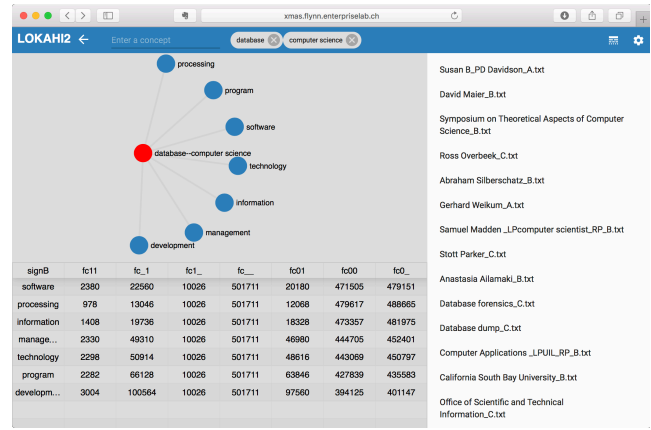

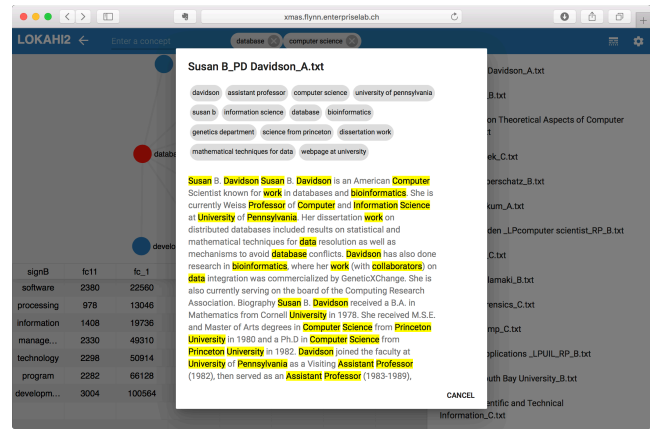

*Figure 2: The Lokahi Prototype: a search engine supported by extracted knowledge graphs.*

As seen in Figure 1 this means that first, entities are recognized and extracted from text. In a second step, relationships between those entities are generated by association learning. And in a third step, by abstraction, these entities and relationships are generalized toward entity sets and relationship sets by classification. The result of this procedure is an automatically extracted semantic data model based on possibly huge amounts of unstructured data. This model represents the essence of the information contained in a collection of data.

The project Lokahi (Kaufmann, Wilke, Portmann, & Hinkelmann, 2014) (Wilke, Emmenegger, Lutz, & Kaufmann, 2016) approached this vision by proposing a method and implementing a system that can extract entities and relationships in a first, basic way. The project XMAS extended on these ideas and developed the prototype further toward concept recognition and n-gram concept extraction (Waldis, Mazzola, & Kaufmann, 2018) In the following, the resulting method and system for entity relationship extraction is presented, and the insights, implications and points for further research are discussed.

## The Lokahi Prototype for Concept Browsing

Lokahi is a research prototype that prototypically explores the automatic generation of knowledge networks. With the Lokahi prototype, automatically tagged texts can be searched and with the help of graph visualization related terms and key phrases can be browsed. The text documents are automatically tagged based on term statistics. The relationships of the individual terms are determined by their common distribution in the corpus. These relationships are then visualized in the Lokahi search engine. As shown in Figure 2, the user can enter search terms to find documents and to browse a knowledge network related to the search query. The interface is designed so that the user can click on the nodes that have a relationship with the concept they are looking for. This allows the user to explore related concepts, to surf in the concept map in the sense of (Nilsson & Palmér, 1999); and to find documents related to concepts. In the example in Figure 2, there are two search terms, "database" and "computer science". The user is displayed a list of documents relevant to this query as well as a concept graph visualizing semantically related concepts to the search query. Clicking on a concept changes the search query term. Clicking on a document shows the content of it, together with extracted key phrases that are also highlighted in the text.

## Extraction of Entities with TF*IDF

For the extraction of entities from text, keyword extraction using a term frequency and inversed document frequency TF*IDF (Lee & Kim, 2008) was chosen as an initial method. For every document d, the terms t are ranked using the TF*IDF score S(t,d) as shown in formula 1, where TF(t,d) is the number of occurrences of t in d, and IDF(t) is the inverse document frequency defined in formula (2), where n is the number of documents in the corpus, and DF(t) is the document frequency of t defined as the number of documents in the index that contain t.

$$S(t,d) = TF(t,d) * IDF(t) \qquad (1)$$

$$IDF(t) = 1 + \log( n / (DF(t) + 1) ) \qquad (2)$$

In later stages, this formula was slightly adapted. Firstly, a variant of the TF*IDF function defined in formula (3) showed better results.

$$S'(t,d) = ( TF(t,d)^2 + IDF(t) ) / |d| \qquad (3)$$

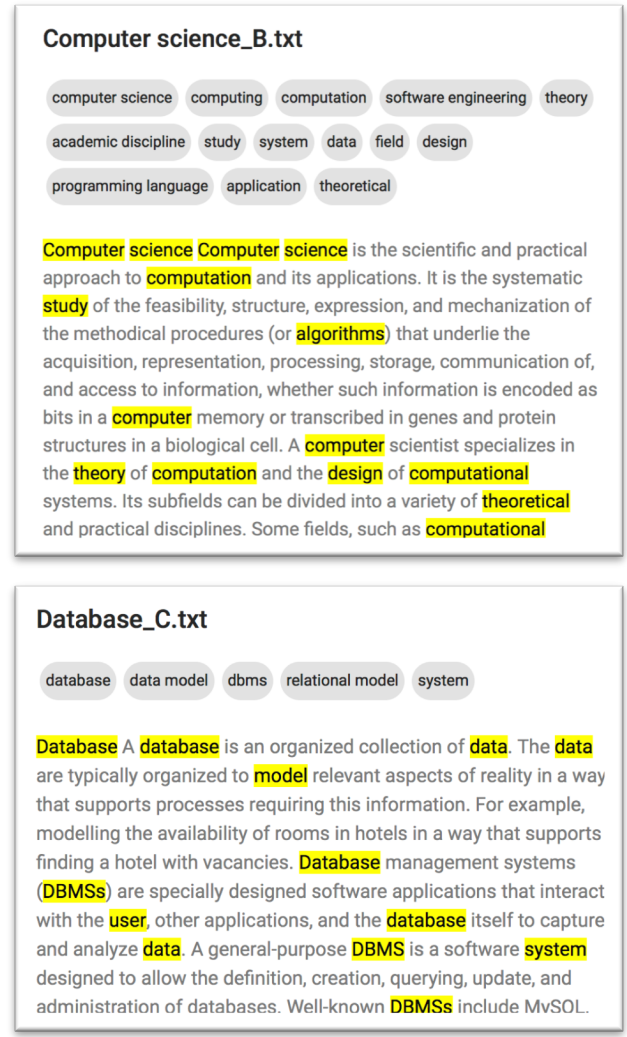

*Figure 3: Keyword extraction for the Wikipedia articles "Computer Science" and "Database"*

In Formula 3, the TF component was squared, and the score was divided by |d|, the number of words in the document to compensate for large TF values in large documents. Also, we implemented a method for combining keywords to n-grams.

This approach was implemented in the Lokahi Prototype by extending the Lucene library source code and by indexing 500K Wikipedia articles of quality levels FA (featured articles), GA (good articles), A, B and C to remove noise from the corpus. In Figure 3 two screenshots of the results of our prototype implementation for key phrase extraction are shown. It is evident the TF-IDF measure can not only match documents based on keywords, but also extract keywords from documents. Also, it is evident that keywords with a high TF-IDF score have a high likelihood to be actual semantic entities.

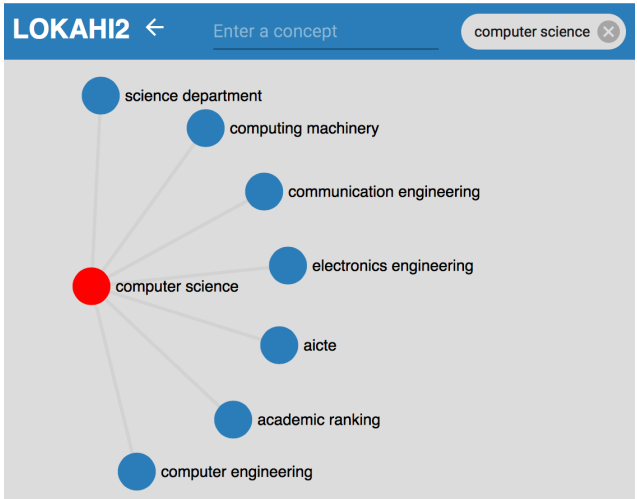
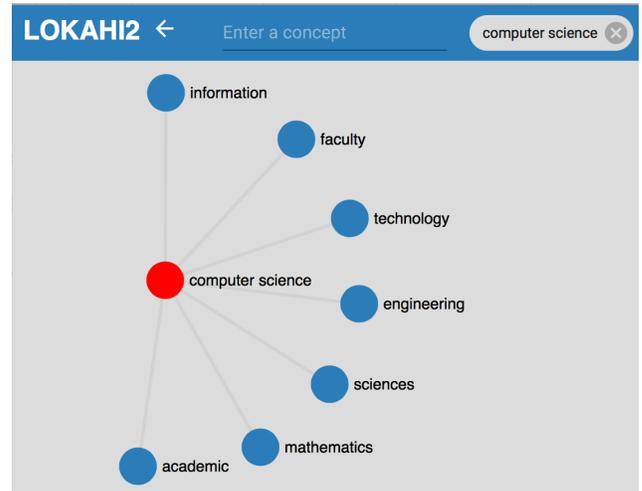
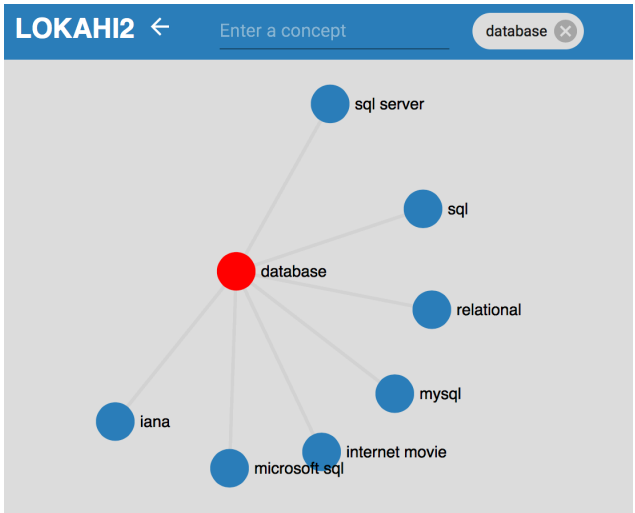
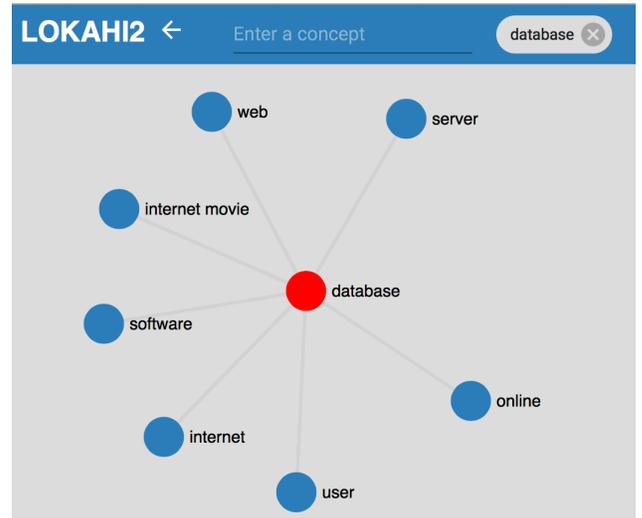

*Figure 4: Extraction of semantic relationships for two terms using likelihood ratios LR*

*Figure 5: Extraction of semantic relationships using pointwise mutual information PMI*

## Extraction of Relationships by Co-Occurrence

As a first step toward relationship extraction, a frequentist approach based on word co-occurrence statistics was chosen as suggested by (Bullinaria & Levy, 2012). Based on the joint probability p(A,B) of document-level co-occurrence of terms A and B, several probabilistic relatedness measures can be computed. Using a frequent itemset approach (Agrawal, Imieliński, & Swami, 1993) the joint frequency of the most frequent keywords in the index can be calculated efficiently. Based on this approach, in the Lokahi prototype, several measures were explored. Two measures turned out most interesting: pointwise mutual information *PMI*, defined in formula 4, and likelihood ratio *LR*, in formula 5.

PMI(a,b) = log( p(a,b) / ( p(a) * p(b) ) )     (4)

LR(a,b) = p(a | b) / p(a | not b)     (5)

This approach was implemented and visualized in a GUI so that the different approaches can be compared qualitatively. In Figure 5, relationships between terms extracted using the PMI measure are visualized for two terms, computer science and database. The related terms in the graph are selected as the top seven items in the ranked list of term pairs according to the PMI measure. Clearly, there is some form of semantic relationship extracted here, because the terms have a similar meaning. In comparison, in Figure 4 the related terms for the same base terms have been computed using the LR measure. Again, there seems to be a semantic similarity between the extracted terms. However, in this case the LR apparently extracts relationships to more specific terms.

## Conclusions and Outlook

The Lokahi prototype demonstrates technologically that it is feasible to extract some form of entities and relationships from text. It is important not to reinvent the wheel, and other research has also already demonstrated this potential that this research confirms. Still, it rather is remarkable that the purely statistical, syntactic computation of Lokahi evidently can extract semantically meaningful entities and relationships. However, this research also indicates the formidable effort needed to tackle the challenge to fulfill the vision of automatic extraction of complete entity relationship models. We can conclude from it some insights and lessons learned.

Firstly, the presented prototype explores the research direction case-based, qualitatively and prototypically. The Lokahi Prototype is a very small first step in this direction. It could be, however, useful for semantic and explorative analysis of unstructured data, for example, in social media mining, if it is extended so that it can easily visualize the semantic structure of any document collection given as input. Secondly, to strengthen the research focus, a broad range of measures for relevance and relatedness ranking needs to be qualitatively and quantitatively assessed. It is important to know what different kinds of semantics different statistics generate. Thirdly, more research into methods to combine single terms to meaningful n-gram entities need to be developed. Perhaps a window- or sentence-based co-occurrence statistic could be compared. Fourthly, even with an optimal extraction of entities and relationships, there needs to be research on methods to automatically classify entities to classes and relationships to association types to form actual entity relationship models. Considering this, we are still a very long way from entity relationship extraction. And fifthly, there is the possibility to incorporate human knowledge, as described by Kaufmann et al. (2014) and Wilke et al. (2016), not only in the form of externalized entities, relationships, classes and associations, e.g. from DBpedia or semantic domain models, but also POS tagging, expert systems and other forms of encoded descriptive and procedural knowledge.

## Acknowledgement

This research has been funded by the Swiss Commission for Technology and Innovation (CTI) as part of the research projects LOKAHI Inside, CTI-No. 16152.1 PFES-ES, and Feasibility Study X-MAS: Cross-Platform Mediation, Association and Search Engine, CTI-No. 26335.1 PFES-ES.

## References

Agrawal, R., Imieliński, T., & Swami, A. (1993). Mining Association Rules Between Sets of Items in Large Databases. In Proceedings of the 1993 ACM SIGMOD International Conference on Management of Data (pp. 207–216). New York, NY, USA: ACM. https://doi.org/10.1145/170035.170072

Alani, H., Kim, S., Millard, D. E., Weal, M. J., Hall, W., Lewis, P. H., & Shadbolt, N. R. (2003). Automatic Ontology-based Knowledge Extraction from Web Documents. Intelligent Systems, 14–21.

Böhm, K., Heyer, G., Quasthoff, U., & Wolff, C. (2002). Topic Map Generation Using Text Mining. Journal of Universal Computing, 8(6). Retrieved from http://www.jucs.org/jucs_8_6/topic_map_generation_using

Bullinaria, J. A., & Levy, J. P. (2012). Extracting semantic representations from word co-occurrence statistics: stop-lists, stemming, and SVD. Behavior Research Methods, 44(3), 890–907. https://doi.org/10.3758/s13428-011-0183-8

Chen, P. P.-S. (1976). The Entity-relationship Model - Toward a Unified View of Data. ACM Trans. Database Syst., 1(1), 9–36. https://doi.org/10.1145/320434.320440

Kaufmann, M., Wilke, G., Portmann, E., & Hinkelmann, K. (2014). Combining Bottom-Up and Top-Down Generation of Interactive Knowledge Maps for Enterprise Search. In R. Buchmann, C. V. Kifor, & J. Yu (Eds.), Knowledge Science, Engineering and Management (pp. 186–197). Springer International Publishing. Retrieved from http://link.springer.com/chapter/10.1007/978-3-319-12096-6_17

Lee, S., & Kim, H. (2008). News Keyword Extraction for Topic Tracking. In 2008 Fourth International Conference on Networked Computing and Advanced Information Management (Vol. 2, pp. 554–559). https://doi.org/10.1109/NCM.2008.199

Maedche, A., & Staab, S. (2001). Ontology learning for the Semantic Web. IEEE Intelligent Systems, 16(2), 72–79. https://doi.org/10.1109/5254.920602

Nilsson, M., & Palmér, M. (1999). Conzilla - Towards a Concept Browser (No. CID-53, TRITA-NA-D9911). Stockholm: Centre for User Oriented IT Design, Dept. Computing Science, Royal Institute of Technology KTH.

Novak, J. D., & Gowin, D. B. (1984). Learning How to Learn. Cambridge University Press.

Quillian, M. R. (1967). Word concepts: A theory and simulation of some basic semantic capabilities. Behavioral Science, 12(5), 410–430. https://doi.org/10.1002/bs.3830120511

Rath, H. H., & Pepper, S. (1999). Topic-Maps: Introduction and Allegro. In Conference Proceedings of the Markup Technologies 99. Philadelphia, USA.

Sowa, J. F. (1976). Conceptual Graphs for a Data Base Interface. IBM Journal of Research and Development, 20(4), 336–357. https://doi.org/10.1147/rd.204.0336

Villalon, J., & Calvo, R. A. (2009). Concept Extraction from Student Essays, Towards Concept Map Mining. In 2009 Ninth IEEE International Conference on Advanced Learning Technologies (pp. 221–225). https://doi.org/10.1109/ICALT.2009.215

Waldis, A., Mazzola, L., & Kaufmann, M. (2018). Concept Extraction with Convolutional Neural Networks (pp. 118–129). Presented at the 7th International Conference on Data Science, Technology and Applications. Retrieved from http://www.scitepress.org/PublicationsDetail.aspx?ID=N8Q5cEQ/jYE=&t=1

Wilke, G., Emmenegger, S., Lutz, J., & Kaufmann, M. (2016). Merging Bottom-Up and Top-Down Knowledge Graphs for Intuitive Knowledge Browsing. In T. Andreasen, H. Christiansen, J. Kacprzyk, H. Larsen, G. Pasi, O. Pivert, … S. Zadrożny (Eds.), Flexible Query Answering Systems 2015 (pp. 445–459). Springer International Publishing. https://doi.org/10.1007/978-3-319-26154-6_34